\documentclass[%
 reprint,
 amsmath,amssymb,
 aps,
 superscriptaddress,
 prd,
]{revtex4-2}

\usepackage[colorlinks,
           bookmarks=true,
           linkcolor=blue,
           urlcolor=blue, 
            anchorcolor=black,
            citecolor=blue
            ]{hyperref}

\usepackage{multirow}
\usepackage{graphicx}% Include figure files
\usepackage{dcolumn}% Align table columns on decimal point
\usepackage{bm}% bold math
\usepackage{float}
\usepackage{exscale}
\usepackage{relsize}

\begin{document}

%\preprint{APS/123-QED}

\title{Energy dependence of $J/\psi$ production in pp collisions with the PACIAE model}% Force line breaks with \\

%\author{Chao Zhang}
%\affiliation{Key Laboratory of Quark and Lepton Physics (MOE) and Institute 
%of Particle Physics, Central China Normal University, Wuhan 430079, China}
%\affiliation{Department of Physics, East Carolina University, 
%  Greenville, North Carolina 27858, USA} 
%\author{Liang Zheng}
%\affiliation{School of Mathematics and Physics, China University of
%  Geosciences (Wuhan), Wuhan 430074, China}
%\author{Shusu Shi} 
%\affiliation{Key Laboratory of Quark and Lepton Physics (MOE) and Institute
%of Particle Physics, Central China Normal University, Wuhan 430079, China}
%\author{Zi-Wei Lin}\email{linz@ecu.edu}
%\affiliation{Department of Physics, East Carolina University, 
 % Greenville, North Carolina 27858, USA} 
 
\author{Kai-Fan Ye}
\affiliation{School of Physics and Information Technology, Shaanxi Normal University, Xi'an 710119, China}
\affiliation{Key Laboratory of Quark and Lepton Physics (MOE) and Institute of Particle Physics, Central China Normal University, Wuhan 430079, China}
\author{Qiang Wang}
\affiliation{School of Physics and Information Technology, Shaanxi Normal University, Xi'an 710119, China}
\author{Jia-Hao Shi}
\affiliation{School of Physics and Information Technology, Shaanxi Normal University, Xi'an 710119, China}
\author{Zhi-Ying Qin}
\affiliation{School of Physics and Information Technology, Shaanxi Normal University, Xi'an 710119, China}
\author{Wen-Chao Zhang}
\email{wenchao.zhang@snnu.edu.cn}
\affiliation{School of Physics and Information Technology, Shaanxi Normal University, Xi'an 710119, China}
\author{An-Ke Lei}
\affiliation{Key Laboratory of Quark and Lepton Physics (MOE) and Institute of
            Particle Physics, Central China Normal University, Wuhan 430079,
            China}
\author{Zhi-Lei She}            
\affiliation{Wuhan Textile University, Wuhan 430200,
            China}
\author{Yu-Liang Yan}
\affiliation{China Institute of Atomic Energy, P. O. Box 275 (10), Beijing
            102413, China}
\author{Ben-Hao Sa}
\affiliation{China Institute of Atomic Energy, P. O. Box 275 (10), Beijing
            102413, China}
\date{\today}% It is always \today, today,

\begin{abstract}

In this work we investigate the $J/\psi$ production in proton-proton  collisions at the center-of-mass energy ($\sqrt{s}$)  equal to 2.76, 5.02, 7, 8 and 13 TeV with a parton and hadron cascade model PACIAE 2.2a. It is based on PYTHIA but extended considering the partonic and hadronic rescatterings before and after hadronization, respectively. In the PYTHIA sector the $J/\psi$ production quantum chromodynamics  processes are selected specially and a bias factor is proposed correspondingly. The calculated total cross sections, the differential cross sections as a function of the transverse momentum  and the rapidity of  $J/\psi$  in the forward rapidity region reproduce the corresponding experimental measurements reasonably well. In the mid-rapidity region,  the double differential cross sections at $\sqrt{s}=$ 5.02, 7 and 13 TeV are also in a good agreement with the experimental data. Moreover, we interpolate the double differential cross section as well as the total cross section of $J/\psi$ in the mid-rapidity region at $\sqrt{s}=$ 8 TeV, which could be validated if the experimental data is available. 

\end{abstract}

\maketitle

\section{\label{sec:intro}Introduction}
  $J/\psi$  is the lightest vector charmonium meson. The suppression of $J/\psi$ production was proposed as a probe to investigate the hot and dense medium, named quark-gluon plasma (QGP),  created in ultra-relativistic nucleus-nucleus collisions \cite{probe_1}. The  $J/\psi$ production could also be suppressed due to the cold nuclear matter effects, such as modifications of nuclear parton distribution functions \cite{cme_1, cme_2}. In order to disentangle the hot and cold medium effects, it is necessary to understand  the $J/\psi$ production in proton-proton (pp) collisions where the initial state effects are absent.

The hadronic $J/\psi$ production mainly results from the gluon-gluon scattering into a $c\bar{c}$ pair in hard scattering and related initial- and final-state radiations, which is described with the perturbative Quantum Chromodynamics (pQCD). The hadronization of $c$ and $\bar{c}$ pair into $J/\psi$ is a soft process, which cannot be dealt with pQCD. The $J/\psi$ production was extensively investigated at colliders such as the Tevatron \cite{Tev_1, Tev_2, Tev_3, Tev_4}, RHIC \cite{Rhic_1} and LHC \cite{lhc_1,lhc_2,lhc_3,lhc_5,lhc_6,lhc_7,lhc_8,lhc_9}. Several theoretical approaches, such as the color singlet model \cite{csm_model}, the non-relativistic QCD model \cite{NRQCD_1, NRQCD_2} and the color evaporation model \cite{CEM_1,CEM_2} have been utilized to describe the experimental data. They differ mostly in the treatment of non-perturbative evolution of the $c\bar{c}$ pair into the bound state $J/\psi$. However, none of these models could simultaneously describe the polarization, the transverse momentum ($p_{\rm T}$) spectrum and the energy dependence of cross sections for $J/\psi$ \cite{lhc_6}. The $J/\psi$ production was also investigated by Monte Carlo simulations. For example, in Refs. \cite{mc_1, mc_2}, the $J/\psi$ production as a function of charged particle multiplicity in pp collisions at the center-of-mass energy ($\sqrt{s}$) equal to 7 and 13 TeV was investigated in the mid-rapidity region by PYTHIA 6.4 \cite{pythia_6} and 8.2 \cite{pythia_8}, respectively. It was found that PYTHIA 8.2 could depict the correlation between the $J/\psi$ yield and the charged particle multiplicity ($dN_{\rm ch}/d \eta$) while PYTHIA 6.4 can not do it well. This could be due to the reason that the multiparton interaction is described differently between them. In Ref. \cite{mc_2}, the correlation between the $J/\psi$ yield and $dN_{\rm ch}/d \eta$ was also explored by EPOS3 event generator \cite{epos_1, epos_2}. In Ref. \cite{Urqmd_1}, the authors utilized a modified ultra-relativistic quantum molecular dynamics (UrQMD) transport model \cite{Urqmd_2, Urqmd_3} to study the $J/\psi$ suppression in high-multiplicity pp collisions at $\sqrt{s}=$ 7 TeV.
 
 Apart from the study in Refs. \cite{mc_1, mc_2,epos_1, pythia_6, pythia_8, Urqmd_1}, we use a parton and hadron cascade model PACIAE 2.2a \cite{paciae} without considering the polarization to investigate the $J/\psi$ production in pp collisions at $\sqrt{s}=$  2.76, 5.02, 7, 8 and 13 TeV.   In the PACIAE model, the $J/\psi$ production QCD processes will be selected specially and a bias factor will be introduced for the simulation sample correspondingly.  The calculated total cross sections, the differential cross sections as a function of the transverse momentum  and the rapidity of  $J/\psi$  in the  forward rapidity region will be, respectively, compared to the corresponding experimental measurements. Meanwhile, at mid-rapidity the total and double differential cross sections  at $\sqrt{s}=$ 5.02, 7 and 13 TeV will also be compared with the experimental data. Moreover, the total and double differential cross sections of $J/\psi$ at $\sqrt{s}=$ 8 TeV will be interpolated in the mid-rapidity region. 

This paper is organized as follows. In sect. \ref{sec:model}, we briefly introduce the PACIAE model. In sect. \ref{sec:method}, we describe the method to produce the simulation sample and to compare the sample with experimental data. In sect. \ref{sec:results}, results and discussions are presented. Finally, the conclusion is given in sect. \ref{sec:conclusions}.

\section{\label{sec:model}The PACIAE model }
\begin{figure*}
\includegraphics[scale=0.39]{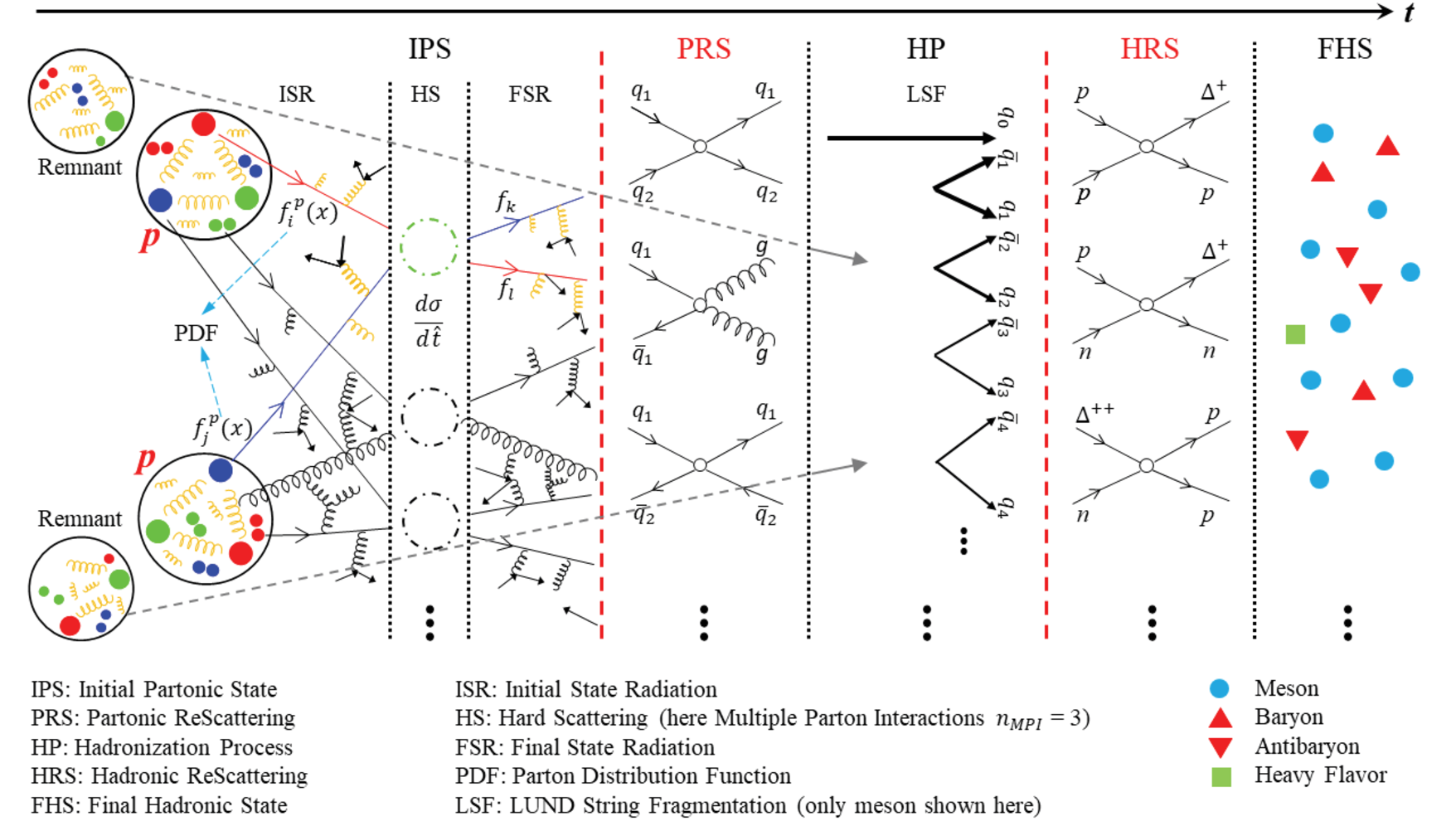}% Here is how to import EPS art
\caption{\label{fig:physical_routine} (Colour online) A sketch of the physical routines in high energy pp collisions \cite{phyrou}.}
\end{figure*}

The PACIAE 2.2a model is designed for elementary collisions. It is based on PYTHIA 6.4 but further considers the partonic rescattering before hadronization and the hadronic rescattering after hadronization. It divides the ultra-relativistic energies pp collisions into four stages: parton initiation, parton rescattering, hadronization, and hadron rescattering. Fig. \ref{fig:physical_routine} is a sketch of the physical routines in a high energy pp collision.

In the first stage, the initial partonic states are created by the hard scatterings, the initial- and final-state radiations in PYTHIA with temporarily switching off the string fragmentation, breaking down the strings and splitting up the diquarks (anti-diquarks) randomly. This partonic matter then undergoes parton rescatterings, where the leading order (LO) pQCD parton-parton interaction cross sections \cite{cs_1, cs_2} are  employed. A $K$ factor is introduced to consider higher order effects and non-perturbative corrections for LO-pQCD parton–parton differential cross sections. After the parton rescattering, the partonic matter is converted into hadrons by the string fragmentation \cite{pythia_6} or the coalescence model \cite{phyrou}.  

 Then followed is the hadronic rescattering. Based on the hadron list after hadronization, two loops over $i$ and $j$ cycling through all hadrons are implemented. If the minimum approaching distance $D$ between two straight line trajectories of $i$ and $j$ particles satisfies $D\leq \sqrt{\sigma^{\rm tot}_{ij}/\pi}$, where $\sigma^{\rm tot}_{ij}$ is the total cross section of these two particles, the particles $i$ and $j$ may collide and the collision time $t_{ij}$ is evaluated \cite{collision_time}. With the $t_{ij}$ of all $i$-$j$ pairs, the initial hadron-hadron collision time list is constructed. A  collision with the least time is picked up from the list and implemented probably.  The hadron list and collision time list are then updated\cite{phyrou}. With the repeat of these two steps until the collision time list is empty,  the kinetic freeze-out happens. For the $J/\psi$ hadronic rescattering, besides the elastic scattering, so far the following inelastic  processes are considered \cite{collision_time, Jpsi_rescattering}:
\begin{eqnarray*}
J/\psi+n\rightarrow \Lambda_c^+ +D^-, \hspace{0.4cm}
J/\psi+n\rightarrow \Sigma_c^+ +D^-, \\[0.8mm]
J/\psi+n\rightarrow \Sigma_c^0 +\bar{D}^0, \hspace{0.4cm}
J/\psi+p\rightarrow \Lambda_c^+ +\bar{D}^0,\\[0.8mm]
J/\psi+p\rightarrow \Sigma_c^+ +\bar{D}^0, \hspace{0.4cm}
J/\psi+p\rightarrow \Sigma_c^{++} +D^-,\\[0.8mm]
J/\psi+\pi^+\rightarrow D^+ +\bar{D}^{*0}, \hspace{0.4cm}
J/\psi+\pi^-\rightarrow D^0 +D^{*-},\\[0.8mm]
J/\psi+\pi^0\rightarrow D^0 +\bar{D}^{*0}, \hspace{0.4cm}
J/\psi+\pi^0\rightarrow D^+ +D^{*-},\\[0.8mm]
J/\psi+\rho^{+}\rightarrow D^+ +\bar{D}^{0}, \hspace{0.4cm}
J/\psi+\rho^-\rightarrow D^0 +D^{-},\\[0.8mm]
J/\psi+\rho^0\rightarrow D^0 +\bar{D}^{0}, \hspace{0.4cm}
J/\psi+\rho^0\rightarrow D^+ +D^{-}.\\
\end{eqnarray*}
There are other models, e.g. the comover interaction model, etc., that also include  hadronic rescatterings \cite{comover_1,comover_2,comover_3, comover_4, comover_5, comover_51, comover_6,  comover_8, comover_81, comover_9, comover_10}. 

\section{\label{sec:method}The method}
The ALICE collaboration had published the inclusive $J/\psi$ production in the forward rapidity region with  $2.5<y<4$ in pp collisions at $\sqrt{s}=$  2.76, 5.02, 7, 8 and 13 TeV in Refs. \cite{lhc_1,lhc_5,lhc_6,lhc_7,lhc_9}. The collaboration also presented the inclusive $J/\psi$ productions in the mid-rapidity region with $-0.9<y<0.9$ in pp collisions at $\sqrt{s}=$  5.02, 7 and 13 TeV in Refs. \cite{lhc_3,lhc_5,lhc_8}. They are listed in Table \ref{tab:dataRef}. The inclusive $J/\psi$ yield contains a prompt component, which includes the direct $J/\psi$ production and the feed-down contribution from the decay of heavier charmonium states, as well as a non-prompt component from the weak decay of beauty hadrons.

\begin{table}[h]
\caption{\label{tab:dataRef}List of ALICE measurements for $J/\psi$ productions.}
\begin{tabular}{c|lc|cc}
\hline\hline
\multicolumn{1}{c}{\multirow{2}{*}{$\sqrt{s}$}} & \multicolumn{2}{c|}{forward-rapidity}                        & \multicolumn{2}{c}{mid-rapidity}                                                    \\ \cline{2-5} 
\multicolumn{1}{c}{}                       & \multicolumn{1}{c|}{interval}                    & reference & \multicolumn{1}{c|}{interval}                       & \multicolumn{1}{l}{reference} \\ \hline
2.76 TeV                                     & \multicolumn{1}{c|}{$2.5<y<4$} &  \cite{lhc_1}   & \multicolumn{1}{c|}{--}                              & --                              \\ \hline
5.02 TeV                                     & \multicolumn{1}{l|}{$2.5<y<4$} &  \cite{lhc_9}  & \multicolumn{1}{c|}{$-0.9<y<0.9$} & \cite{lhc_3}                              \\ \hline
7 TeV                                        & \multicolumn{1}{l|}{$2.5<y<4$} & \cite{lhc_6}         & \multicolumn{1}{c|}{$-0.9<y<0.9$} & \cite{lhc_5}                              \\ \hline
8 TeV                                        & \multicolumn{1}{l|}{$2.5<y<4$} & \cite{lhc_7}         & \multicolumn{1}{c|}{-- }                              & --                              \\ \hline
13 TeV                                       & \multicolumn{1}{l|}{$2.5<y<4$} & \cite{lhc_9}         & \multicolumn{1}{c|}{$-0.9<y<0.9$} &  \cite{lhc_9}                                 \\ \hline
\hline
\end{tabular}
\end{table}

\begin{table*}
\caption{\label{tab:sigma_of_chane} Summary of cross sections ($\sigma_{\rm sel}$) for specially selected $J/\psi$ `color-singlet' process in pp collisions at various energies from PACIAE simulation. The last row lists the total $J/\psi$ cross-section in one event at a given energy. This cross section can be calculated by the weighted sum of the cross sections for the  $J/\psi$ selection process $\sigma_{\rm sel}$, with the weight taken as the product of the fraction of heavier charmonium state's radiatively decaying into $J/\psi$ (i.e. branching ratio (BR)) and the sampling  probability.} 
\begin{tabular}
{c@{\hspace{0.2cm}}|c@{\hspace{0.2cm}}c@{\hspace{0.2cm}}c@{\hspace{0.2cm}}c@{\hspace{0.2cm}}c@{\hspace{0.2cm}}|c@{\hspace{0.2cm}}|c}
\hline\hline

\multicolumn{1}{c|}{\multirow{2}{*}{Processes}}                         & 2.76 TeV   & 5.02 TeV   & 7 TeV   & 8 TeV   & 13 TeV   &     \multirow{2}{*}{BR}                &          \multirow{2}{*}{Sampling probablity}                            \\ \cline{2-6}
\multicolumn{1}{c|}{} & \multicolumn{5}{c|}{$\sigma_{\rm sel}(\mathrm{\mu b})$}          &  &  \\ \cline{1-8}

$gg\rightarrow   J/\psi g$                    & 32.871   & 62.654   & 64.798  & 83.623  & 132.548  & 100.0\%             & 2.184\%                               \\
$gg\rightarrow \chi_{0c}g$                   & 241.830  & 499.519  & 533.856 & 700.603 & 1169.812 & 1.4\%               & 21.054\%                              \\
$gg\rightarrow \chi_{1c}g$                    & 138.797  & 312.055  & 345.636 & 460.827 & 823.761  & 34.3\%              & 5.000\%                               \\
$gg\rightarrow \chi_{2c}g$                    & 265.777  & 552.470  & 593.993 & 779.302 & 1302.222 & 19.0\%              & 23.101\%                              \\
$gg\rightarrow \chi_{0c}$                     & 294.944  & 554.780  & 572.512 & 744.378 & 1202.169 & 1.4\%               & 23.525\%                             \\
$gg\rightarrow \chi_{2c}$                      & 255.635  & 484.268  & 502.488 & 651.972 & 1060.098 & 19.0\%              & 25.083\%                              \\
$gg\rightarrow J/\psi \gamma$                 & 0.929    & 1.872    & 1.856   & 2.588   & 4.232    & 100.0\%             & 0.053\%                               \\ \hline
                                              %& \multicolumn{5}{c|}{$\sigma_{J/\psi}(\mathrm{\mu b})$} &                     &                                      \\ \cline{2-6}
$\sigma_{J/\psi}(\mathrm{\mu b})$             & 28.631      & 57.349      & 60.822   & 79.524   & 132.111   &           --          &   --    \\
\hline\hline
\end{tabular}
\end{table*}

In this paper, as a complementary work to the study performed by event generators such as PYTHIA 6.4, PYTHIA 8.2, EPOS3 and UrQMD \cite{mc_1,mc_2,Urqmd_1}, we investigate the $J/\psi$ production in pp collisions at different energies with PACIAE 2.2a. In the model, a `menus' of subprocesses for the $J/\psi$ production is composed with the sets of `MSEL=0' and `MSUB($i$)=1, $i$=86, 87, 88, 89, 104,105,106'. This especial setting is completely the same as that in PYTHIA 6.4. It corresponds to the following `color-singlet' processes, $gg\rightarrow J/\psi g$, $gg\rightarrow \chi_{0c} g$, $gg\rightarrow \chi_{1c} g$, $gg\rightarrow \chi_{2c} g$, $gg\rightarrow \chi_{1c}$, $gg\rightarrow \chi_{2c}$, $gg\rightarrow J/\psi \gamma$. The higher mass excited states, like the $\chi_{0c}$, $\chi_{1c}$ and $\chi_{2c}$ may subsequently decay to  $J/\psi$. Since the $\psi(2s)$ production channel is not available in the PYTHIA 6.4 default processes, we did not consider the contribution of $\psi(2s)$ to  $J/\psi$. As shown in Ref. \cite{color_octet}, in the region with $p_{\rm T}>$  7 GeV/c, the $J/\psi$ production may be dominated by the `color-octet’ processes. In this work, we only focus on the production of $J/\psi$ in the low and intermediate $p_{\rm T}$ range below 7 GeV/c. Thus the `color-octet’ processes are not considered. Relative to the default simulation without above extra setting,  the simulated results with extra setting are bias samples. In order to get rid of this bias effect, for example, we rescale the cross sections in simulation with the first moment of the $J/\psi$ $p_{\rm T}$ spectra and  match it with the similarly rescaled cross section in experiment, 
\begin{equation}
\dfrac{\left.\frac{\mathrm{d}\sigma_{J/\psi}}{\mathrm{d}y}\right|_{\mathrm{sim}}}{\mathlarger{\int} p_{\mathrm{T}}\left.\frac{\mathrm{d^2}\sigma_{J/\psi}}{\mathrm{d}p_{\mathrm{T}}\mathrm{d}y}\right|_{\mathrm{sim}} \mathrm{d}p_{\mathrm{T}}}=\dfrac{\left.\frac{\mathrm{d}\sigma_{J/\psi}}{\mathrm{d}y}\right|_{\mathrm{exp}}}{\mathlarger{\int} p_{\mathrm{T}}\left.\frac{\mathrm{d^2}\sigma_{J/\psi}}{\mathrm{d}p_{\mathrm{T}}\mathrm{d}y}\right|_{\mathrm{exp}} \mathrm{d}p_{\mathrm{T}}}.
\end{equation}
Thus, in order to compare the simulated samples with the experimental data, we propose a bias factor, 
 
\begin{equation}
\begin{aligned}
 B &=\dfrac{\mathlarger{\int}  p_{\mathrm{T}}\left.\frac{\mathrm{d^2}\sigma_{J/\psi}}{\mathrm{d}p_{\mathrm{T}}\mathrm{d}y}\right|_{\mathrm{exp}}\mathrm{d}p_{\mathrm{T}}}{\mathlarger{\int} p_{\mathrm{T}}\left.\frac{\mathrm{d^2}\sigma_{J/\psi}}{\mathrm{d}p_{\mathrm{T}}\mathrm{d}y}\right|_{\mathrm{sim}} \mathrm{d}p_{\mathrm{T}}}\\
   & =\dfrac{\mathlarger{\int}  p_{\mathrm{T}}\left.\frac{\mathrm{d^2}\sigma_{J/\psi}}{\mathrm{d}p_{\mathrm{T}}\mathrm{d}y}\right|_{\mathrm{exp}}\mathrm{d}p_{\mathrm{T}}}{\mathlarger{\int} \left.p_{\mathrm{T}}\frac{\mathrm{d^2}N_{J/\psi}}{\mathrm{d}p_{\mathrm{T}}\mathrm{d}y}\cdot \frac{\sigma_{J/\psi}}{N_{J/\psi}}\right|_{\mathrm{sim}} \mathrm{d}p_{\mathrm{T}}}, \\ \label{eq:Rescaledfactor}
\end{aligned}
\end{equation}
 which has to be multiplied to the simulated results. Here we have inserted the relation between the differential yield and differential cross section (see Appendix for detail). In Eq. (\ref{eq:Rescaledfactor}),  $N_{J/\psi}$ is total $J/\psi$ yield in simulation, $\sigma_{J/\psi}$ is total $J/\psi$ cross section in one event returned by PACIAE 2.2a and is tabulated in Table  \ref{tab:sigma_of_chane}.  Similar treatments are applied to  the differential rapidity distribution of $J/\psi$.
%For the differential cross section as a function of $p_{\rm T}$  in the middle and forward rapidity regions, the bias factor $B$ is defined as
% the relation between the differential  $J/\psi$ yield and differential cross section in simulation is derived in detail. 

In PACIAE, the model parameters are chosen as the default values in the PYTHIA 6.4 model, except for the $K$ factor, which is determined by fitting the simulation to the experimental data in pp collisions at a given energy with a least $\chi^{2}s$ method. For the $J/\psi$ differential cross section as a function of $p_{\rm T}$ in the forward rapidity region, the $\chi^{2}$ is defined as 
\begin{equation}
\chi^{2}=\sum_{i=1}^{N}{\frac{(y^i-f(p^i_{\rm T};K))^2}{(\Delta^i)^2}}, \label{eq:Kfactor}
\end{equation}
where $N$ is the number of data points, $y^i$ is the experimental value in the $i^{\rm th}$ bin, $f(p^i_{\rm T};K)$ is the simulated value in the same bin, $\Delta^i$ is the total error of the experimental data. Table \ref{tab:Kfactor} gives the $\chi^{2}$ values divided by the number of degrees of freedom ($\chi^{2}/ndf$) for different $K$ factors at a given energy. The best $K$ factor at that energy is determined by minimizing the  corresponding $\chi^{2}$ values and is shown by boldface type. At a given collision energy, the $K$ factor for the $J/\psi$ production in the mid-rapidity region is assumed to be the same as that in the forward rapidity region.

\begin{table}[h]
\caption{\label{tab:Kfactor}Summary of $\chi^{2}/ndf$ for different $K$ factors at a given energy. The minimum $\chi^{2}/ndf$ is shown by boldface type.}
\begin{tabular}
{c@{\hspace{0.2cm}}|c@{\hspace{0.2cm}}c@{\hspace{0.2cm}}c@{\hspace{0.2cm}}c@{\hspace{0.2cm}}c}
\hline\hline
\multicolumn{1}{c|}{\multirow{2}{*}{$K$}}                         & 2.76 TeV   & 5.02 TeV   & 7 TeV   & 8 TeV   & 13 TeV             \\ \cline{2-6}
\multicolumn{1}{c|}{} & \multicolumn{5}{c}{$\chi^{2}/ndf$}  \\ \cline{1-6}
1.1 & 5.39/6 & 14.06/6 & 23.85/6 & 17.09/6 & 66.34/7  \\
1.2 & 4.74/6 & 10.93/6  & 19.57/6 & 16.77/6 & 71.67/7  \\
1.3 & 6.11/6 & 17.25/6 & \textbf{16.62/6} & 16.40/6 & 63.80/7  \\
1.4 & 4.81/6 & 13.55/6  & 18.87/6 & 17.08/6 & 60.94/7  \\
1.5 & 6.86/6 & 11.44/6  & 21.23/6 & \textbf{13.40/6} & 58.07/7 \\
1.6 & \textbf{3.46/6} & 13.10/6 & 19.18/6 & 13.71/6 & 58.83/7  \\
1.7 & 5.92/6 & \textbf{10.92/6}  & 19.54/6 & 14.85/6 & \textbf{54.86/7}  \\
1.8 & 7.04/6 & 14.89/6 & 19.25/6 & 14.37/6 & 65.33/7  \\
1.9 & 5.96/6 & 15.22/6 & 20.05/6 & 14.94/6 & 71.83/7 \\
\hline\hline
\end{tabular}
\end{table}

\section{\label{sec:results}Results and discussions}

\begin{figure}[h]
\includegraphics[scale=0.4]{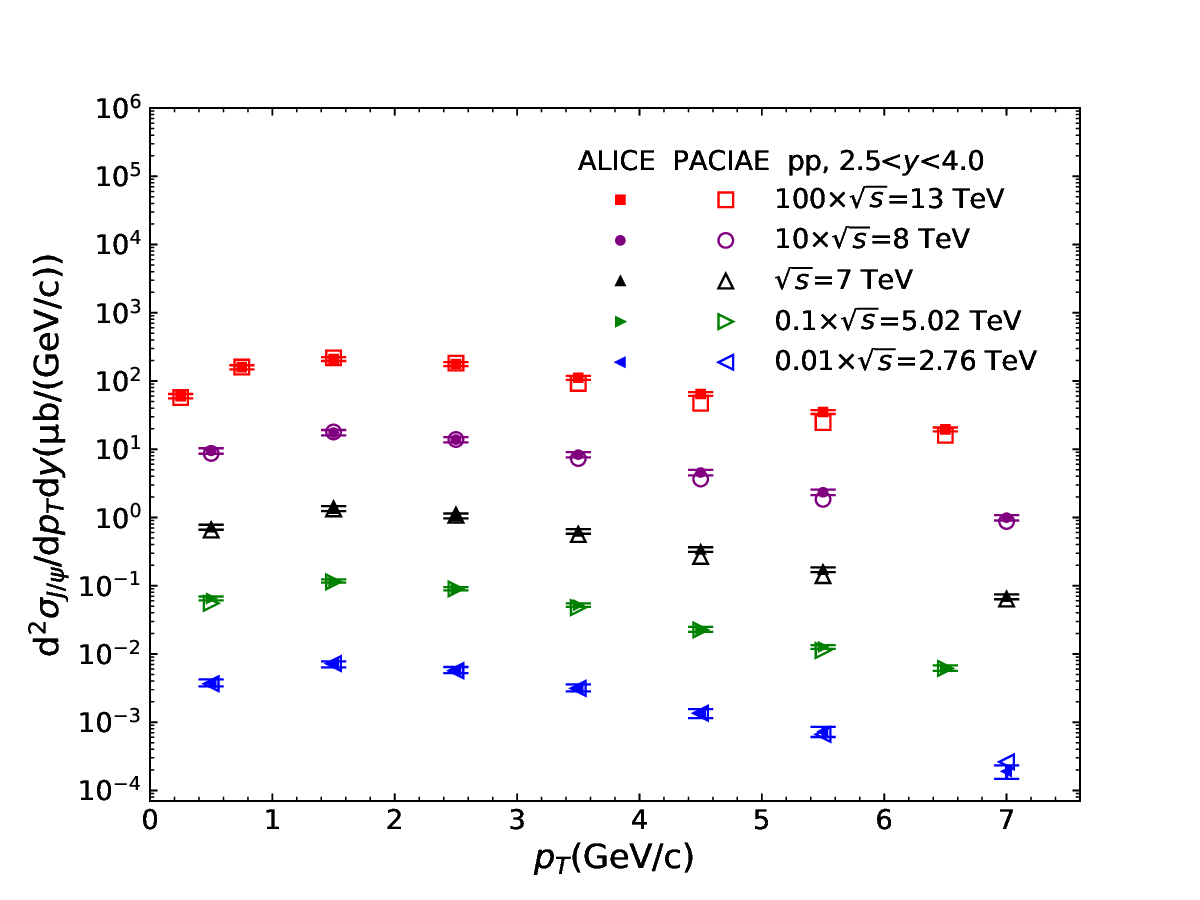}% Here is how to import EPS art
\caption{\label{fig:Jpsi_pt_spectra_forward_region} (Colour online) Inclusive  $J/\psi$ double differential cross sections as a function of $p_{\mathrm{T}}$ at forward rapidity in pp collisions with $\sqrt{s}=$ 2.76, 5.02, 7, 8 and 13 TeV. The solid symbols are experimental data taken from Refs. \cite{lhc_1,lhc_9,lhc_6,lhc_7,lhc_9}. The hollow symbols are the results from the PACIAE model. }
\end{figure}

\begin{figure}[h]
\includegraphics[scale=0.4]{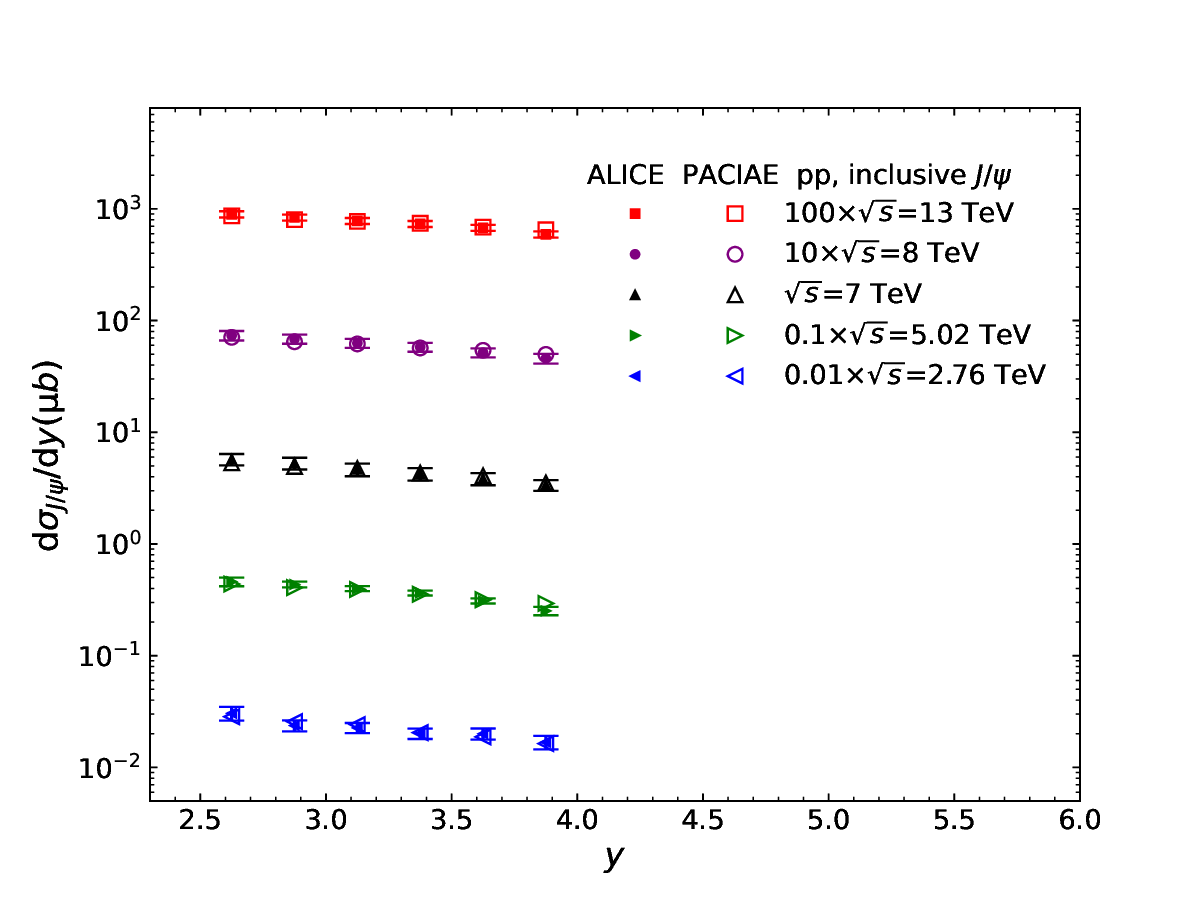}% Here is how to import EPS art
\caption{\label{fig:Jpsi_y_dist_forward_region} (Colour online) Inclusive  $J/\psi$  single differential cross sections as a function of $y$ in pp collisions at $\sqrt{s}=$2.76, 5.02, 7, 8 and 13 TeV. The solid symbols are experimental data taken from Refs. \cite{lhc_1,lhc_9,lhc_6,lhc_7,lhc_9}. The hollow symbols are the results from the PACIAE model.}
\end{figure}

Fig. \ref{fig:Jpsi_pt_spectra_forward_region} shows the inclusive $J/\psi$  cross sections from PACIAE (hollow symbols) at forward rapidity in pp collisions at $\sqrt{s}=$  2.76, 5.02, 7, 8 and 13 TeV, comparing with experimental data (solid symbols). The error bars attached to the data points represent the total uncertainties of the $J/\psi$ cross sections. It is found that the simulations well describe the experimental data. Fig. \ref{fig:Jpsi_y_dist_forward_region} presents the inclusive  $J/\psi$ cross sections as a function of rapidity in the forward region. In order to make the simulations comparable with experimental data, we employed the same $p_{\mathrm{T}}$ cut in the former as those applied in the latter at each energy. It's observed that the simulation well reproduces the corresponding experimental measurement. 

\begin{figure}[h]
\includegraphics[scale=0.4]{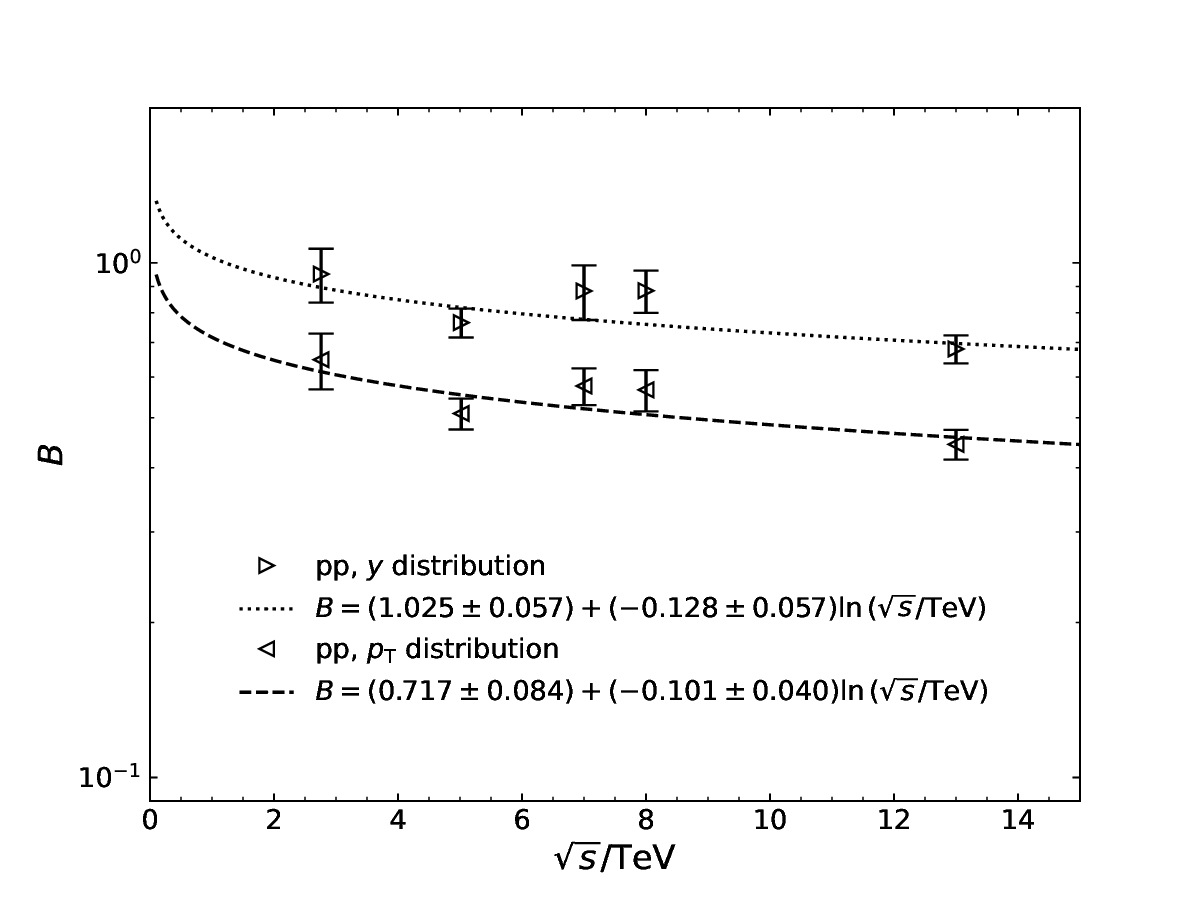}% Here is how to import EPS art
\caption{\label{fig:B_Ecm} 
The energy dependence of the bias factor in the forward region in pp collisions at $\sqrt{s}=$ 2.76, 5.02, 7, 8 and 13 TeV. The left (right) triangles represent the bias factors calculated for the $p_{\mathrm{T}}$ ($y$) distributions. The dash and dotted curves represent the fitted functions.}
\end{figure}

\begin{figure}[h]
\includegraphics[scale=0.3]{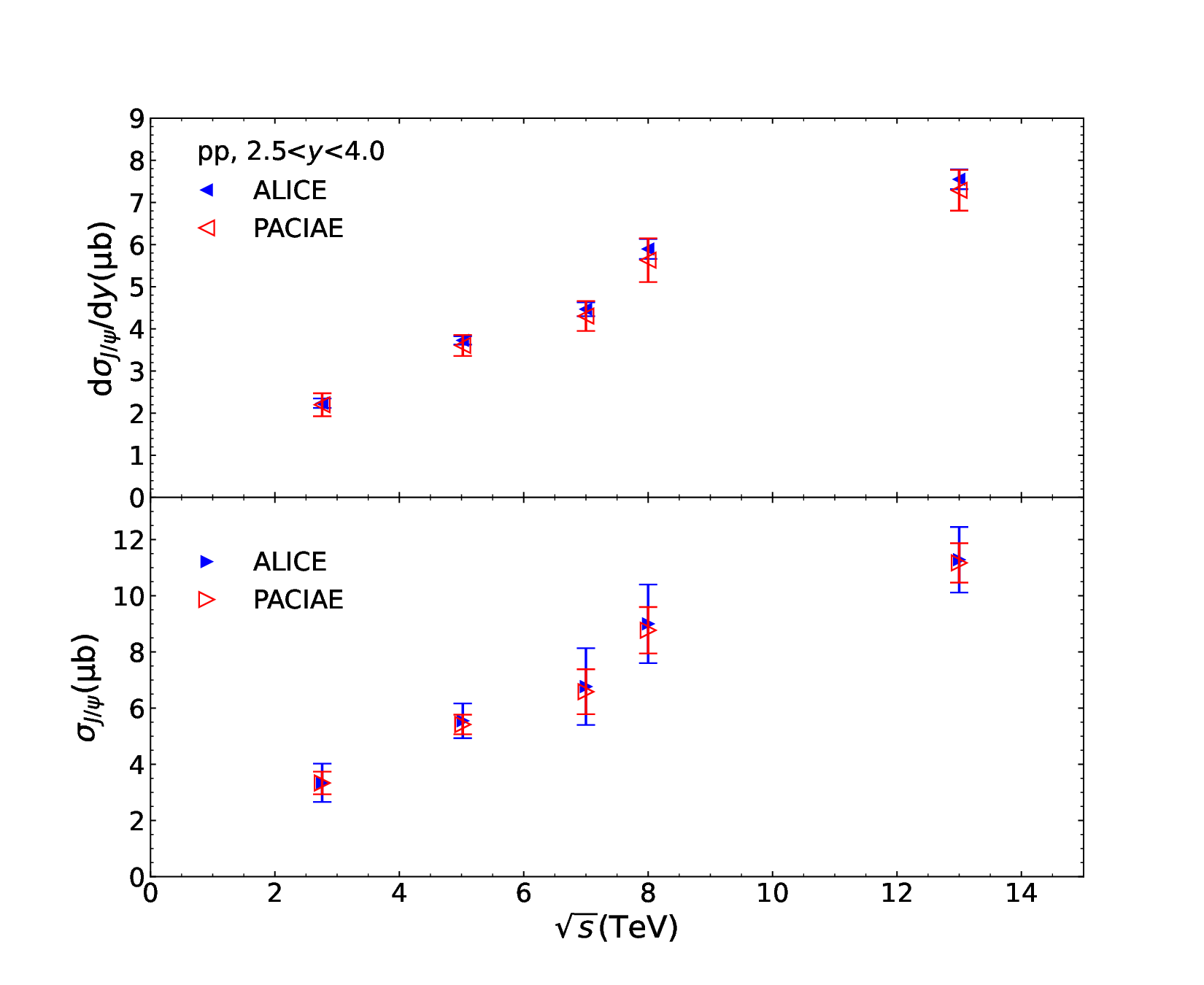}% Here is how to import EPS art
\caption{\label{fig:TotalNum_Ecm} 
 (Colour online) Upper (Lower) panel shows the $J/\psi$ rapidity differential cross section (total cross section) as a function of $\sqrt{s}$ in the forward rapidity region in pp collisions. The solid (hollow) symbols represent the cross sections  from the experimental data (the PACIAE model).}
\end{figure}

As described in sect. \ref{sec:method}, due to  the especial selection of the $J/\psi$ production processes in the simulation, the bias factor is introduced when comparing the simulation with experimental data. The energy dependence of the bias factors introduced, respectively, for the $J/\psi$ $y$ and $p_{\rm T}$ distributions are presented in Fig. \ref{fig:B_Ecm}. It is shown that at a given energy the former one is larger than the latter one. In addition, for both cases the bias factor slightly decreases with energy. In order to guide the eyes, we parameterize the dependence of $B$ on energy with a fit function $B=a+b\textrm{ln}(\sqrt{s}/\textrm{TeV})$, where $a$ and $b$ are free parameters. The fitted curves are also presented in the figure. The upper (lower) panel in Fig. \ref{fig:TotalNum_Ecm} shows the $J/\psi$ rapidity differential  (total) inclusive cross sections as a function of $\sqrt{s}$ in the forward rapidity region. It is observed that the simulations from the PACIAE model are in a good agreement with the experimental data.   Moreover, there is a steady increase of the cross sections with the increasing collision energy. At a given energy, the total cross section is about a factor of 1.5 larger than the rapidity differential one.  The bias factor for the $J/\psi$ rapidity distribution is also 1.5 times larger than that for the $p_{\rm T}$ distribution. This factor is exactly the bin size in the forward rapidity region.

\begin{figure}[h]
\includegraphics[scale=0.4]{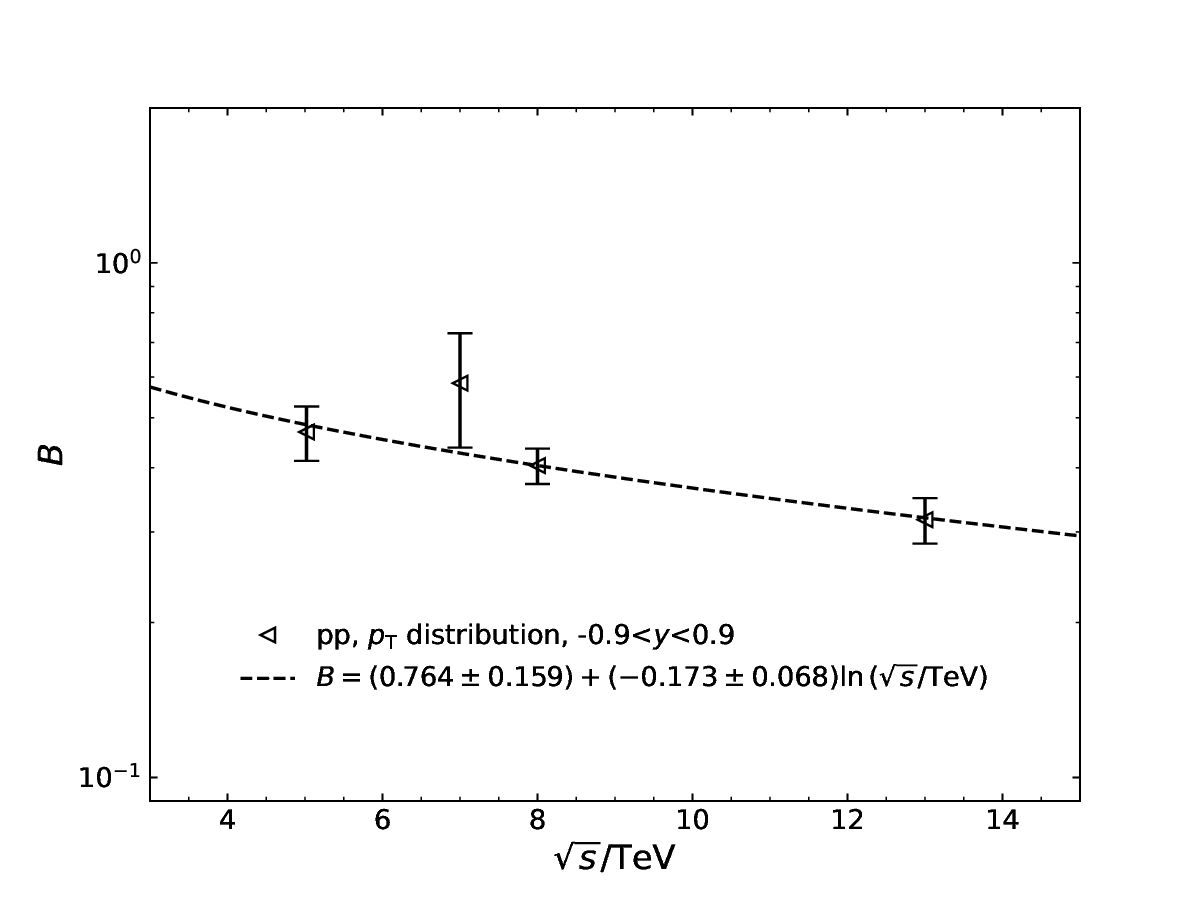}% Here is how to import EPS art
\caption{\label{fig:B_Ecm_midyy} The energy dependence of the bias factor in the mid-rapidity region in pp collisions at $\sqrt{s}=$ 5.02, 7 and 13 TeV. The dash  curve represents the parameterization formular. The bias factor at $\sqrt{s}=$ 8 TeV is  interpolated with this parameterization formular.}
\end{figure}

\begin{figure}[h]
\includegraphics[scale=0.4]{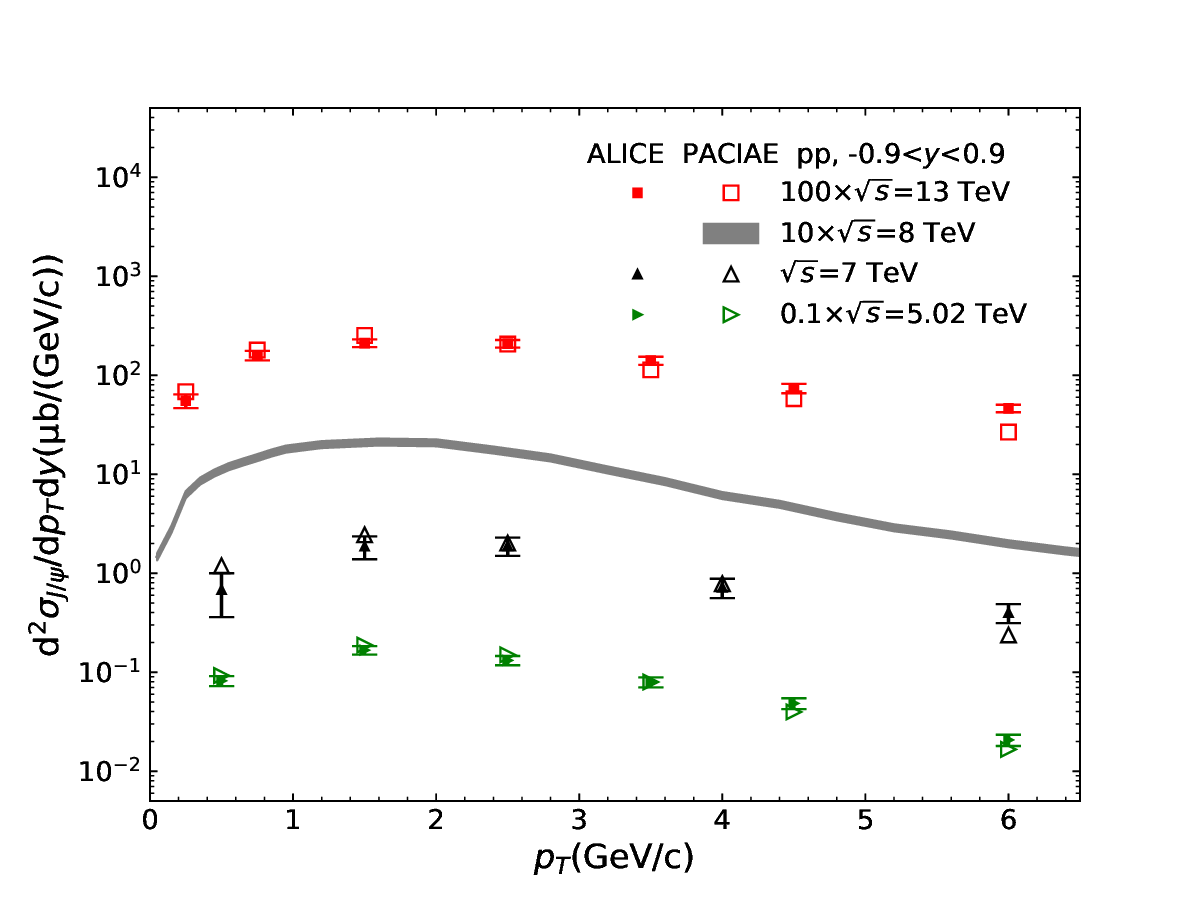}% Here is how to import EPS art
\caption{\label{fig:Plot_DNDpT_midyy} 
 (Colour online) Inclusive  $J/\psi$  double differential cross sections as a function of $p_{\mathrm{T}}$ at mid-rapidity in pp collisions at $\sqrt{s}=$5.02, 7 and 13 TeV. The solid symbols are experimental data taken from Refs. \cite{lhc_3,lhc_5,lhc_9}. The hollow symbols are the results from the PACIAE model. The gray band represents the interpolated results from the PACIAE model in pp collisions at $\sqrt{s}$= 8 TeV.
}
\end{figure}

We use the same method as that in the forward rapidity region to determine the bias factor of the $J/\psi$ double differential cross sections in the mid-rapidity region. Fig. \ref{fig:B_Ecm_midyy} shows the energy dependence of the bias factor at mid-rapidity  in pp collisions at $\sqrt{s}=$5.02, 7 and 13 TeV. This energy dependence can be parameterized as $B=(0.764\pm0.159)+(-0.173\pm0.068)\textrm{ln}(\sqrt{s}/\textrm{TeV})$. The bias factor at $\sqrt{s}=$ 8 TeV is then interpolated with this parameterization formular. It is equal to $0.404\pm0.032$. Fig. \ref{fig:Plot_DNDpT_midyy} presents the inclusive $J/\psi$ double differential cross section as a function of  $p_{\mathrm{T}}$  in the mid-rapidity region of $-0.9<y<0.9$ in pp collisions at $\sqrt{s}=$  5.02, 7 and 13 TeV. It is found that the simulations agree with experimental data reasonably well. 

\begin{figure}[h]
\includegraphics[scale=0.4]{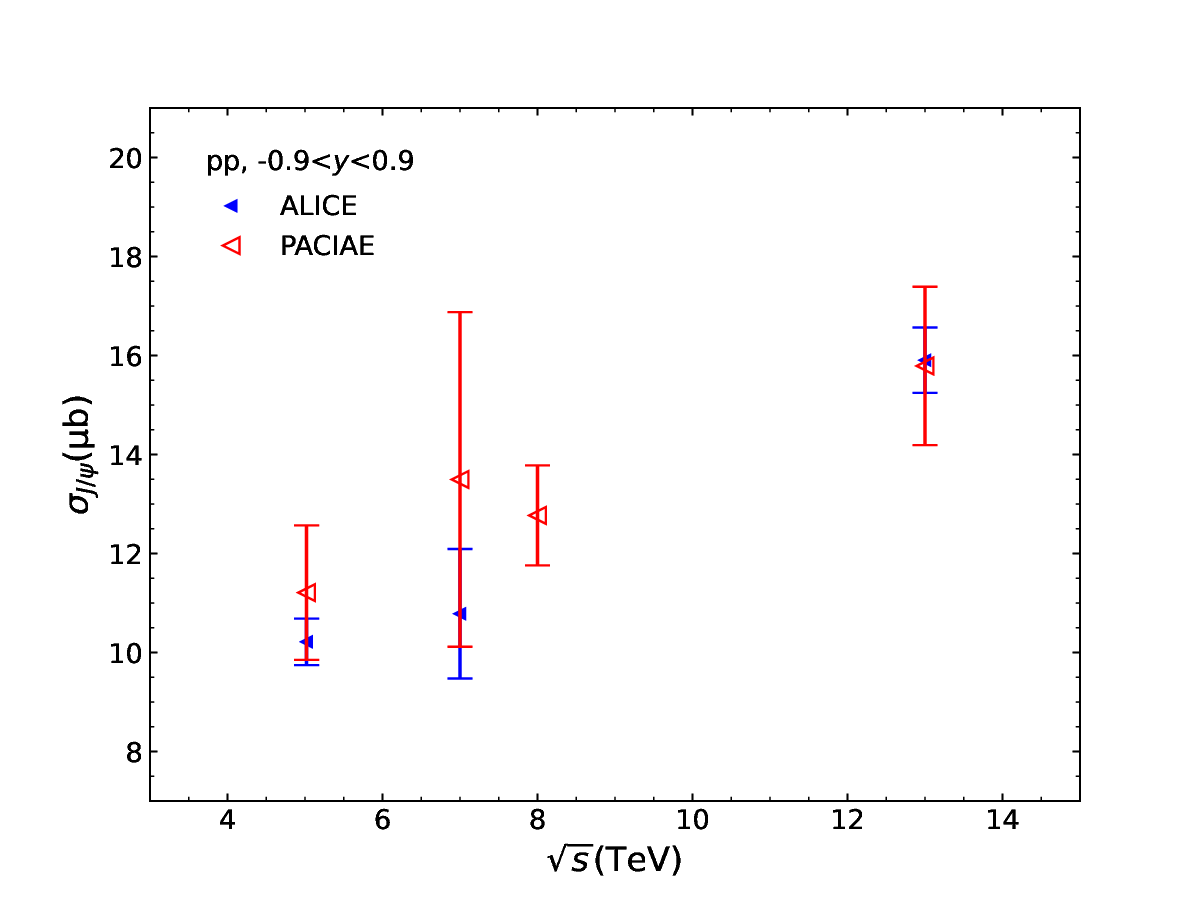}% Here is how to import EPS art
\caption{\label{fig:TotalNum_Ecm_midyy} 
 (Colour online) Total inclusive $J/\psi$ cross section at mid-rapidity in pp collisions with $\sqrt{s}=$ 5.02, 7 and 13 TeV.  The solid (hollow) symbols represent the cross sections  from the experimental data (the PACIAE model). The total cross section at $\sqrt{s}=$ 8 TeV is an interpolation based on the integration of the double differential cross section over the mid-rapidity region.}
\end{figure}

In sect. \ref{sec:method}, we have determined the $K$ factors by fitting the simulated $J/\psi$  $p_{\rm T}$ differential cross sections to the experimental data in the forward rapidity region in pp collisions at $\sqrt{s}=$ 8 TeV. Together with the assumption that  the $K$ factor at mid-rapidity is the same as that in the forward rapidity region (see Table \ref{tab:Kfactor}), we could generate the simulation sample for the $J/\psi$ production in the mid-rapidity region in pp collisions at $\sqrt{s}=$ 8 TeV.  With the application of the bias factor interpolated at $\sqrt{s}=$ 8 TeV (see Fig. \ref{fig:B_Ecm_midyy})  to the simulated sample,  the double differential  $J/\psi$ cross section at mid-rapidity in pp collisions at $\sqrt{s}=$ 8 TeV is available.  It is presented as the gray band in Fig. \ref{fig:Plot_DNDpT_midyy}. Fig. \ref{fig:TotalNum_Ecm_midyy} shows the energy dependence of the simulated total inclusive $J/\psi$ cross section at mid-rapidity in pp collisions at $\sqrt{s}=$ 5.02, 7 and 13 TeV. At a given energy, the total cross section in simulation agrees with experimental data within uncertainties. The inclusive total $J/\psi$ cross section at  $\sqrt{s}=$ 8 TeV is evaluated by integrating the double differential cross section over $p_{\mathrm{T}}$ and $y$. The above interpolated $J/\psi$ double differential  cross section and the total cross section in pp collisions at $\sqrt{s}=$ 8 TeV could be validated if the experimental data is available.

\section{\label{sec:conclusions}Conclusions}
In this paper, we have investigated the $J/\psi$ production in pp collisions at $\sqrt{s}=$ 2.76, 5.02, 7, 8 and 13 TeV with  the parton and hadron cascade model PACIAE 2.2a. It is based on PYTHIA but differs from PYTHIA in the addition of the parton rescattering before hadronization and the hadron rescattering after hadronization.  In the model the $J/\psi$ production QCD processes are selected specially and the bias factor is proposed and applied to the simulation sample correspondingly. The calculated $J/\psi$ total cross section, the differential cross section as a function of $p_{\rm T}$ and $y$ in the forward rapidity region agree with the corresponding experimental measurements reasonably well. In the mid-rapidity region,  the double differential cross sections of  $J/\psi$ at $\sqrt{s}=$ 5.02, 7 and 13 TeV also reproduce the experimental data. Moreover, the $J/\psi$ double differential cross section and  the total cross section at $\sqrt{s}=$ 8 TeV are interpolated. They could be validated if the experimental data is available. 

We have noticed that the ALICE collaboration had published the results of $J/\psi$ polarization in pp collisions at $\sqrt{s}=$ 7 and 8 TeV \cite{polarization_1,polarization_2}. We would like to see whether the PACIAE model could simultaneously describe the $J/\psi$ polarization and  $p_{\rm T}$ spectrum. Moreover, we realize the necessary of the individually investigating the partonic and hadronic rescattering effects. These investigations will be presented in our next works. 

\begin{acknowledgments}

We would like to thank Profs. D.-M. Zhou, X.-M. Zhang and G.-Y. Qin at Central China Normal University, and Prof. Y.-Q. Ma at Peking University for their valuable discussions. This work is supported by the research fund from the School of Physics and Information Technology at Shaanxi Normal University, by the Scientific Research Foundation for the Returned Overseas Chinese Scholars, State Education Ministry, by Natural Science Basic Research Plan in Shaanxi Province of China (program No. 2023-JC-YB-012) and by the National Natural Science Foundation of China under Grant Nos. 11447024, 11505108 and 12375135.
\end{acknowledgments}

 \section*{Appendix}
 \setcounter{equation}{0}
  \setcounter{table}{0}
\setcounter{subsection}{0}
\renewcommand{\theequation}{A\arabic{equation}}
\renewcommand{\thetable}{A\arabic{table}}
\renewcommand{\thesubsection}{A\arabic{subsection}}
In this appendix, we present a detailed derivation of  the relation between the $J/\psi$ differential  yield and differential cross section. The total $J/\psi$ cross section in one event at a given energy in simulation, $\sigma_{J/\psi}$, can be expressed as:
\begin{equation}
    \sigma_{J/\psi}=\sigma_{J/\psi}^{ev}\dfrac{N_{J/\psi}}{N_{ev}}=\sigma_{J/\psi}^{ev}N_{J/\psi}^{per},\label{eq:A1}
\end{equation}
where $N_{J/\psi}$ ($N_{ev}$) is the number of $J/\psi$s (events) in the simulated sample, and $N_{J/\psi}^{per}=N_{J/\psi}/N_{ev}$ is the number of $J/\psi$s in one event. In this work, for each energy, $N_{ev}$ is set to be $10^6$.  $\sigma_{J/\psi}^{ev}=\sigma_{J/\psi}/N_{J/\psi}^{per}$ is the cross section of producing one $J/\psi$ in one event. With Eq.(\ref{eq:A1}), the $J/\psi$ differential cross section  can be expressed  as 
\begin{equation}
    \dfrac{\mathrm{d^2}\sigma_{J/\psi}}{\mathrm{d}p_{\mathrm{T}}\mathrm{d}y}=\dfrac{\sigma_{J/\psi}^{\mathrm{ev}}}{N_{\mathrm{ev}}}\dfrac{\mathrm{d^2}N_{J/\psi}}{\mathrm{d}p_{\mathrm{T}}\mathrm{d}y}=\dfrac{\sigma_{J/\psi}}{N_{J/\psi}}\dfrac{\mathrm{d^2}N_{J/\psi}}{\mathrm{d}p_{T}\mathrm{d}y}.
\end{equation}
i.e. 
\begin{equation}
\dfrac{1}{\sigma_{J/\psi}}\dfrac{\mathrm{d^2}\sigma_{J/\psi}}{\mathrm{d}p_{\mathrm{T}}\mathrm{d}y}=\dfrac{1}{N_{J/\psi}}\dfrac{\mathrm{d^2}N_{J/\psi}}{\mathrm{d}p_{T}\mathrm{d}y}.
\end{equation}
This is exactly the relation between the differential yield and differential cross section used in the simulation.

%\nocite{*}

%\section*{References}

\end{document}